\documentclass[pra]{revtex4}

\usepackage{amsmath}
\usepackage{graphicx}
\begin{document}
\newcommand{\e}{\mathrm{e}}
\newcommand{\x}{{\bf r}}
\newcommand{\K}{{\bf k}}
\newcommand{\y}{{\bf y}}
\newcommand{\D}{{\rm d}}
\newcommand{\p}{{\bf p}}
\newcommand{\tr}{\mathrm {Tr}}
\newcommand{\de}{:=}

\title{Statics and dynamics of BEC's in double square well potentials.}

\author{ E. Infeld$^{1}$, P. Zi\'{n}$^{2}$, J. Goca{\l}ek$^{3}$ and
M. Trippenbach$^{1,2}$}

\affiliation{$^{1}$ Soltan
Institute for Nuclear Studies, Ho\.{z}a 69, PL-00-681 Warsaw, \\
$^2$ Institute for Theoretical Physics, Warsaw University, Ho\.{z}a
69, PL-00-681 Warsaw, \\ $^{3}$Polish Academy of Science, Al.
Lotnik\'{o}w� 32/46, 02-668 Warszawa Poland.}

\begin{abstract}
In this paper we treat the behavior of Bose Einstein condensates in
double square well potentials, both of equal and different depths.
For even depth, symmetry preserving solutions to the relevant
nonlinear Schr\"{o}dinger equation is known, just as in the linear
limit. When the nonlinearity is strong enough, symmetry breaking
solutions also exist, side by side with the symmetric one.
Interestingly, solutions almost entirely localized in one of the
wells are known as an extreme case. Here we outline a method for
obtaining all these solutions for repulsive interactions. The
bifurcation point at which, for critical nonlinearity, the
asymmetric solutions branch off from the symmetry preserving ones is
found analytically. We also find this bifurcation point and treat
the solutions generally via a Josephson Junction model.

When the confining potential is in the form of two wells of
different depth, interesting new phenomena appear. This is true of
both the occurrence of the bifurcation point for the static
solutions, and also of the dynamics of phase and amplitude varying
solutions. Again a generalization of the Josephson model proves
useful. The stability of solutions is treated briefly.
\end{abstract}

\maketitle

\section{Introduction}\label{Introduction}

The nonlinear Schr\"{o}dinger equation is a powerful tool for
describing Bose Einstein condensates at zero temperature. Double
well potentials are an important class of configurations to which
this tool can be applied. For square wells, exact solutions are known
to exist \cite{Reinhardt}. In Zi\'{n} {\it et al.} \cite{Zin}, we
outlined a method for obtaining such exact solutions for a symmetric
double square well situation with attractive interaction. We found
an exact criterion to determine the bifurcation point. Here we
perform similar calculations for the repulsive case and extend our
treatment, including wells of different depth but also a stability
analysis. The repulsive case is perhaps more interesting in view of
the fact that, for this interaction, situations such that most of
the condensate was contained in one of the wells have been seen
experimentally \cite{Oberthaler}. We also present some dynamic
calculations not included in \cite{Zin} for both kinds of
interaction. These lean somewhat on a generalized Josephson Junction
model \cite{Adams,elena,Giovanazzi,Smerzi,Raghavan}. We note in
passing that a Josephson Junction for a Bose-Einstein condensate was
first obtained by Inguscio's group \cite{JosBEC}, see also
\cite{Oberthaler}.

Symmetry breaking solutions that are known to exist for positive
nonlinearity (repulsive interaction in the case of BEC, dark
solitons in nonlinear optical media)  often tend to localize the
wave function in one of the wells. This happens for the nonlinearity
exceeding a critical value at which the asymmetric solutions branch
off from the symmetry preserving ones in the parameter space.
Therefore, we can talk about bifurcation at this critical value of
nonlinearity. The existence of such solutions of the nonlinear
Schr\"{o}dinger equation was first pointed out in the context of
molecular states for repulsive interaction \cite{Davis}, as will be
treated here. Importantly, the effect of this spontaneous symmetry
breaking has been observed in photonic lattices \cite{kevrekidis}.
It should be stressed that the nature of bifurcation depends on the
symmetry of the problem and is of the pitchfork variety for even
wells and saddle point for uneven wells \cite{Ioos}.

In this paper we will consider a double square well potential, first
symmetric and then asymmetric. The asymmetric potential leads to
more complicated profiles. As far as we know, these square well
configurations are the only ones for which exact solutions exist.
These solutions are all in the form of Jacobi elliptic functions.
One of the problems considered here, extending \cite{Zin} to
repulsive interaction, is how to proceed from easily obtainable
symmetric double well solutions of the linear Schr\"{o}dinger
equation to the fully nonlinear case, and from so obtained symmetric
solutions on to the bifurcated, asymmetric ones. When the {\em
potential} is asymmetric both bifurcation of the static solutions
and the dynamics of oscillating solutions will be seen to become
very different from those for the symmetric potential.

The manuscript is composed as follows: In section \ref{sec1} we
derive symmetry preserving states starting from the linear limit and
then gradually increasing the nonlinear interaction. In section III
we investigate the symmetry breaking states that branch off from the
symmetry preserving ones in the parameter space. We give a simple
exact formula for the bifurcation point. Section IV treats
asymmetric potentials. Section V is devoted to dynamics treated by
the Josephson model, particularly useful at the bifurcation point,
and then numerically. Results are consistent by all three methods
(sections III, IV and V).  Some concluding remarks wind up the text
(section VI). Heavier calculations have been relegated to the
Appendix.

This paper can be read independently of reference \cite{Zin}.

\section{Antisymmetric states from the linear limit (symmetric wells)}\label{sec1}

We start from the nonlinear Schr\"{o}dinger equation
\begin{equation}\label{NLS}
\left[-\frac{\partial^2}{\partial x^2} + V(x) + \eta |f(x)|^2
\right] f(x)=\mu
 f(x)
\end{equation}

\begin{figure}
\includegraphics[width=8.5cm]{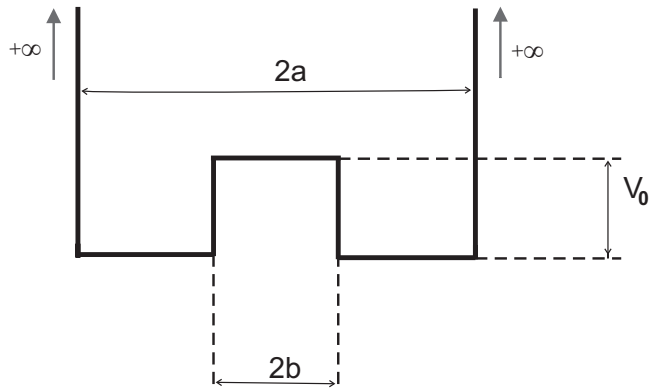}
\caption{Symmetric double square well potential. In the following $2a =1$, $2b=0.1$ and $V_0=300$.} \label{rys1}
\end{figure}

Here the potential is of the form
\begin{equation}\label{one}
V(x) = \left\{\begin{array}{lll} V_0 & \mbox{ for } |x| \leq b \\
0 & \mbox{ for } b < |x| \leq a\\
\infty & \mbox{ for } |x| \geq a
\end{array} \right.
\end{equation}
See Fig.~\ref{rys1}. Solutions in the three regions will be
written as
\begin{equation}\label{two}
f(x) = \left\{\begin{array}{lll} f_1(x) & \mbox{ for } -a \leq x < -b \\
f_2(x) & \mbox{ for } |x| \leq b\\
f_3(x) & \mbox{ for } b < x \leq a
\end{array} \right.
\end{equation}
The solutions vanish on and outside the outer boundaries $|x|\ge a$. We
assume continuity of $f(x)$ and its derivative at $x=\pm b$   and
normalization to $\int^a_{-a}|f(x)|^2\,dx=1$. The symmetric
solutions are
\begin{eqnarray}\label{symetric}
f_1(x) &=&  A\,\, \mbox{sn}(k(x+a)|m)
\\ \nonumber
f_2(x) &=&  A_2\,\,  \mbox{nc}(k_2x|m_2)
\\ \nonumber
f_3(x) &=& -A\,\, \mbox{sn}(k(x-a)|m)
\end{eqnarray}
and the antisymmetric solutions, which will be of particular interest
here, are
\begin{eqnarray}\label{antisymetric}
f_1(x) &=&  A\,\, \mbox{sn}(k(x+a)|m)
\\ \nonumber
f_2(x) &=&  -A_2\,\, \mbox{sc}(k_2x|m_2)
\\ \nonumber
f_3(x) &=& A\,\, \mbox{sn}(k(x-a)|m).
\end{eqnarray}
Here $f_1(x)$ and $f_3(x)$ have been chosen to be zero at the ends,
and also so as to preserve even and odd parity respectively for the
two cases. The parameters of the symmetric solutions are found from
Eq.~(\ref{NLS}) to satisfy ($V_0 > \mu$)
\begin{eqnarray} \label{eq1}
A^2 = \frac{2m k^2}{\eta}, \ \ \ \ \ \   A_2^2 =
\frac{2(1-m_2)k_2^2}{\eta}, \ \ \ \ \ \ 
 \mu = (1+m)k^2 = (1-2m_2)k_2^2 + V_0,
\end{eqnarray}
and for the antisymmetric solutions we have
\begin{eqnarray} \label{eq2}
A^2 = \frac{2m k^2}{\eta}, \ \ \ \ \ \   A_2^2 =
\frac{2(1-m_2)k_2^2}{\eta},  \ \ \ \ \ \
\mu = (1+m)k^2 = (m_2-2)k_2^2 + V_0.
\end{eqnarray}
Positive
roots for all the A's are taken throughout. We choose $\mu$, $m$ and $m_2$ to generate all
the other constants.  These three parameters  will
determine the solution completely.

We now concentrate on the antisymmetric case, as we have checked
that bifurcation only occurs for this case in the lowest mode as
suggested by Fig. \ref{rys3} (and also by Fig. 3 of Ref. \cite{Dagosta}). We have two continuity
conditions at $\pm b$
\begin{eqnarray}
&&A \mbox{sn}(k\omega|m) = -A_2 \mbox{sc}(-k_2b|m_2) \label{8}
\\
&&A k\mbox{cn}(k\omega|m)\mbox{dn}(k\omega|m) = -A_2 k_2
\mbox{dc}(-k_2b|m_2)\mbox{nc}(-k_2b|m_2). \label{9}
\end{eqnarray}
Here $\omega = a-b$. The normalization of the wave function,
$\int_{-a}^{a} \mbox{d}x \, |f(x)|^2 =1 $ yields
\begin{equation}\nonumber
2A^2 \int_a^b   \mbox{sn}^2(k(x-a)|m) \mbox{d}x + 2A_2^2 \int_0^b
\mbox{sc}^2(k_2x|m_2) \mbox{d} x =1
\end{equation}
The above normalization condition works out as:
\begin{eqnarray}\label{warnorm}
{4k_2} \left[ \mbox{sn}(k_2b|m_2)\
\mbox{dc}(k_2b|m_2)-E(k_2b|m_2)\right] + {4k}\left[ k\omega -
 E(k \omega |m)
\right] ={\eta},
\end{eqnarray}
where $E(u|m)$ is the elliptic function of the second kind
\cite{Abram}. We now have three equations for $m$, $m_2$ and $\mu$
as required. Here $a$, $b$, $V_0$ and $\eta$ are fixed and describe
one specific experimental setup. Up to now, the shooting method was
used to find solutions \cite{Reinhardt}.

To systematically solve our equations we first turn to the linear
limit $\eta=0$, $k^2=\mu$, $k_2^2=V_0-\mu$. The functions are now
easy to calculate
\begin{eqnarray} \nonumber
f_1(x) &=&  A\,\,\sin k(x+a)
\\ \nonumber
f_2(x) &=& - A_2\,\,  \sinh k_2x
\\ \nonumber
f_3(x) &=& A\,\, \sin k(x-a).
\end{eqnarray}
The two continuity conditions are:
\begin{eqnarray}\label{wargranlin}
A\,\, \sin k\omega &=&  A_2\sinh k_2b
\\
Ak\,\, \cos k\omega &=& -A_2 k_2\,\, \cosh k_2b ,
\label{wargranlin2},
\end{eqnarray}
This last equation will give us linear $\mu$ in terms of fixed
parameters. The normalization condition is
\begin{eqnarray}\label{eq5}
A^2 = \left( \omega - \frac{\sin 2 k\omega }{2 k} + \frac{\sin^2
k\omega }{ \sinh^2 k_2b } \left( -b + \frac{\sinh 2 k_2b}{2 k_2}
\right) \right)^{-1}.
\end{eqnarray}
Giving a value for $A$, and $A_2$ follows from continuity. We are
now ready to tie this solution up to the small $\eta$ limit in a
perturbative manner. We notice that $A, A_2, k, k_2$ obtained in the
linear approach become a zero order approximation in an $\eta$
expansion. The parameters $m$ and $1-m_2$ are both of order $\eta$
and follow from equations (\ref{eq1}) and (\ref{wargranlin})
\begin{equation}\label{eq6}
m = \frac{A^2}{2 k^2}\eta \ \ \ \ 1-m_2 = \frac{\sin^2 k\omega }{
\sinh^2 k_2b } \frac{A^2}{2 k_2^2}\eta.
\end{equation}
  Linear $\mu$, denoted by
$\mu_0$, is found as the lowest root of
\begin{equation}\label{eq7}
\sqrt{\mu_0}\cot(\sqrt{\mu_0}\omega)+\sqrt{V_0-\mu_0}\coth(\sqrt{V_0-\mu_0}b)=0.
\end{equation}
There will also be a small $\eta$ correction $\Delta \mu$ such that
$\mu = \mu_0+\Delta \mu$. This will complete the calculation of the
three unknowns $\mu$, $m$, $m_2$ in the small $\eta$ limit
(Appendix).

Now that we have a starting point, we can generate all symmetric
solutions by gradually increasing $\eta$. We introduce the notation
$\mu \equiv m_0, \,\,\, m \equiv m_1$, and $m_2$. We write the
conditions (\ref{8}) and (\ref{9})  in the symbolic functional form
\begin{eqnarray}\label{newcondit1}
h_0(m_0,m_1,m_2) = k\sqrt{m_1}\,\,
\mbox{sn}(k\omega|m_1)-k_2\sqrt{(1-m_2)}\,\, \mbox{sc}(k_2b|m_2)=0,
\\ \label{newcondit2}
h_1(m_0,m_1,m_2) = k^2\sqrt{m_1}\,\,
\mbox{cn}(k\omega|m_1)\,\,\mbox{dn}(k\omega|m_1)+k_2^2\sqrt{1-m_2}\,\,
\mbox{dc}(k_2b|m_2)\,\,\mbox{nc}(k_2b|m_2)=0.
\end{eqnarray}
Here we used Eq.~(\ref{eq1}) to express the amplitudes $A$ and $A_2$
in terms of the $m_i$ and the wavevectors $k$ and $k_2$, which in
turn can be expressed in terms of the $m_i$. The left hand side of
equation (\ref{warnorm}) defines $h_2(m_0,m_1,m_2)$, which is
evidently free of $\eta$. Upon defining $\eta_0=0,\, \eta_1=0,\,
\eta_2=\eta$ we write all three conditions (\ref{8}), (\ref{9}) and
(\ref{warnorm}) simply as
\begin{equation}\label{eqh}
h_i(m_0,m_1,m_2)=\eta_i \ \ \ \mbox{for} \ \ \ i = 0,1,2.
\end{equation}
In all three equations (\ref{eqh}) functions $h_i(m_0,m_1,m_2)$ on
the left, remain free of $\eta$. Hence if we increase $\eta$ by a
small increment $\Delta \eta$ the parameters $m_i$ will increase by
$\Delta m_i$ governed by
\begin{equation}\label{increment}
\left(\frac{\partial h_i}{\partial m_j}\right) \Delta m_j = \Delta
\eta_i,
\end{equation}
where $\Delta \eta_i=(0,0,\Delta \eta)$.  Assuming the matrix
$\left(\frac{\partial h_i}{\partial m_j}\right)$ to be nonsingular
we can now generate increments in $m_i$ by gradually increasing the
control parameter $\eta$. Inverting Eq.~(\ref{increment}) we find
\begin{equation}\label{decrement}
\Delta m_i = \left(\frac{\partial h_i}{\partial m_j}\right)^{-1}
\Delta \eta_j.
\end{equation}

\section{Symmetry Breaking States (symmetric wells)}\label{sec2}

Even when the double well is symmetric, the nonlinear
Schr\"{o}dinger equation is known to admit symmetry breaking states.
This is in contradistinction to the linear version, admitting only
symmetric and antisymmetric states as treated in section
(\ref{sec1}). These symmetry breaking states are possible above a
critical value of $\eta$.  They
are of considerable physical interest, as they include situations
such as the location of most of the wavefunction in one half of the
double well. More generally, there is the possibility of very
different profiles in the two halves. The solutions corresponding to
symmetry breaking are known to bifurcate from the antisymmetric
ones. Here we will give a condition defining the bifurcation points
in parameter space and investigate how this bifurcation can be
interpreted. We will give diagrams to illustrate this. Similar
diagrams for a quartic potential can be found in \cite{Dagosta},
however they do not correspond to any analytic solutions known to
us.

Solutions generalizing the symmetric case are:
\begin{eqnarray} \nonumber
f_1(x) &=&  A_1\,\, \mbox{sn}(k_1(x+a)|m_1)
\\ \nonumber
f_2(x) &=&  A_2\,\,  \mbox{nc}(k_2(x+d)|m_2)
\\ \nonumber
f_3(x) &=& -A_3\,\, \mbox{sn}(k_3(x-a)|m_3)
\end{eqnarray}
and the generalization for the antisymmetric case is:
\begin{eqnarray} \nonumber
f_1(x) &=&  A_1\,\, \mbox{sn}(k_1(x+a)|m_1)
\\ \nonumber
f_2(x) &=&  -A_2\,\, \mbox{sc}(k_2(x+d)|m_2)
\\ \nonumber
f_3(x) &=& A_3\,\, \mbox{sn}(k_3(x-a)|m_3).
\end{eqnarray}
When $d=0$, $m_1=m_3=m$ the solutions (\ref{symetric}) and
(\ref{antisymetric}) are recovered. Once again we concentrate on a
generalization of the antisymmetric case, as the only one branching
off from a basic mode.

We now have five conditions for $\mu, m_1, m_2, m_3, d$, which we
will denote $m_I$ \  \ $I=0,... 4$ and in place of equation
(\ref{eqh}) we have $g_I=\eta_I$, see the Appendix. One solution is
$d=0, m_1=m_2=m$, as we know, and conditions (\ref{8}), (\ref{9})
and (\ref{warnorm}) are recovered. However, as we will see, above a
certain threshold in $\eta$ a second solution appears. The value of
$\eta$ at which this bifurcation occurs will be denoted by
$\eta_{bif}$. The second solution branches off the antisymmetric one
at this point. To find it we note that at such a point the
antisymmetric solution is continuous with respect to $\eta$, whereas
the asymmetric one is not. Therefore we expect the $3 \times 3$
matrix $\left(\frac{\partial h_i}{\partial m_j}\right)$ to be
nonsingular, whereas the $5 \times 5$ matrix $\left(\frac{\partial
g_I}{\partial m_J}\right)$ will be singular at this point. Simple
algebra shows that the determinant of the $5 \times 5$ matrix can be
factorized at the bifurcation point for which $m_1 = m_3$ and $d=0$.
\begin{eqnarray}
\det \left( \frac{\partial g_I}{\partial m_J} \right) = \det \left(
\frac{\partial h_i}{\partial m_j} \right) D_2
\end{eqnarray}
and $D_2$ is found to be given by:
\begin{eqnarray}
D_2 = 2 \left[\frac{\partial g_0}{\partial m_1}\frac{\partial
g_2}{\partial d}  -\frac{\partial g_0}{\partial d} \frac{\partial
g_2}{\partial m_1} \right]_{m_1=m_3, d=0}
\end{eqnarray}
In view of the above, $D_2 = 0$. This condition can be expressed in
terms  of variables {\em characterizing the antisymmetric solution}.
If we write conditions (\ref{newcondit1}), (\ref{newcondit2}) as
$h_0 = h_{0}^{(1)} - h_{0}^{(2)}$ and $h_1 = h_1^{(1)} + h_1^{(2)}$
we obtain a simple condition for the bifurcation point:
\begin{eqnarray}\label{newcond2}
\partial_b h_1^{(2)} \partial_m h_0^{(1)} + \partial_m h_1^{(1)}\partial_b
h_0^{(2)} = 0.
\end{eqnarray}
This further simplifies to:
\begin{equation}
\frac{\partial}{\partial m} \left[ mk^4 - mk^2V_0
\mbox{sn}^2(k\omega |m)     \right] = 0.\label{nieglupi}
\end{equation}

\begin{figure}
\includegraphics[width=8.5cm]{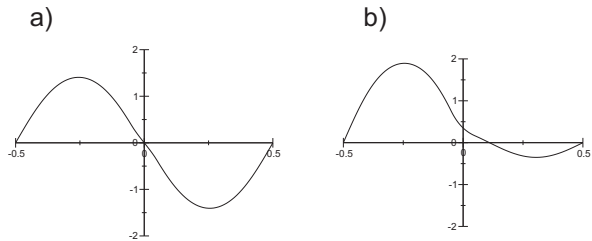}
\caption{Antisymmetric and asymmetric solutions for the same value
of $\eta=10$.} \label{rysstany}
\end{figure}

Thus we can find the bifurcation point on the antisymmetric branch
in terms of just two of the antisymmetric variables. Having that
point we can move out onto the asymmetric branch using:
\begin{equation}\label{brak}
\Delta m_I = \left(\frac{\partial g_I}{\partial m_J}\right)^{-1}
\Delta \eta_J,  \ \ \ \ \ \ \eta_I = 0 \ \ \ \mbox{for} \ \ \
I=0,1,2,3 \ \ \ \mbox{and} \ \ \  \eta_4=\eta.
\end{equation}
This equation can be used everywhere except at the branch point,
where second  derivatives must come in. We have checked numerically
that condition (\ref{nieglupi}) is always satisfied at the
bifurcation point. For an illustration of a bifurcated asymmetric
solution paired with the corresponding symmetry preserving one far from
bifurcation, see Fig.~\ref{rysstany}.

\section{Asymmetric Potential}

Suppose now that the right hand well is somewhat shallower that 
on the left ($V=V_3, \,\, b<x<a, \,\, 0 < V_3 < \mu$). Otherwise we keep the notation
of Fig.~\ref{rys1}. Eqn.~(\ref{eq1}) is now replaced by
\begin{equation} \label{eq1n}
A_1^2 = \frac{2m_1 k_1^2}{\eta}, \ \ \   A_2^2 =
\frac{2(1-m_2)k_2^2}{\eta},  \ \ \ A_3^2 = \frac{2m_3 k_3^2}{\eta},
\ \ \   \mu = (1+m_1)k_1^2 = (1-2m_2)k_2^2 + V_0=(1+m_3)k_3^2+V_3.
\end{equation}
Equation (\ref{eq2}) is similarly modified. When this is done all
formulas for the symmetry breaking case formally  carry through, with the
understanding that $k_3^2$ is now $(\mu-V_3)/(1+m_3)$. Interchanging $m_1$ and $m_3$ no longer gives a trivial alteration. Solutions with most of the condensate on the left or on the right 
are no longer mirror images. Also, the
linear limit is altered, relevant equations becoming
\begin{eqnarray} \nonumber
f_1(x) &=&  A_1\,\,\sin k_1(x+a)
\\ \nonumber
f_2(x) &=& - A_2\,\,  \sinh k_2(x+d)
\\ \nonumber
f_3(x) &=& A_3\,\, \sin k_3(x-a).
\end{eqnarray}
Even in this limit, we now have five equations for five unknowns:
$m_1$, $m_2$, $m_3$, $\mu$ and $d$. This limit is thus nolonger
much simpler than the fully nonlinear case. However this is not
worth pursuing, as the interesting bifurcation does not now occur
from the "linear" branch.

Illustrations of how phase diagrams are modified as compared to the
symmetric potential case are given in Fig.~\ref{rys2}, E and F. As
we increase $\eta$ from zero, a new double fixed point suddenly
appears and bifurcates as we increase $\eta$ (see the next section).
Thus we have two new fixed points (nolonger a pitchfork
bifurcation).

\section{A Josephson Junction approach}

Now allow $f$ to be time dependent and satisfy the one dimensional
equation
\begin{equation}\label{gpt}
i \frac{\partial f(x,t)}{\partial t}  =
\left[-\frac{\partial^2}{\partial x^2} + V(x) + \eta |f(x,t)|^2
\right] f(x,t),
\end{equation}
with potential $V(x)$ in the form of a double well, which is not
necessarily symmetric. To establish a link between the above equation
and the Josephson model, we first focus on the energy spectrum of
the system in the linear limit ($\eta =0$). It consists of pairs
of energy levels separated by a gap that is proportional to the height of
the barrier; for a sufficiently high barrier the spacing between the
pairs is larger than the spacing within the first
pair. In this case we can construct a variational analysis based
on the lowest pair of levels, $\psi_1$ and $\psi_2$. We assume that
$f(x,t)$ is normalized to unity and approximate it by $f(x,t) \simeq
a_{L}(t)w_{L}(x)+a_{R}(t)w_{R}(x)$, where $w_{L,R}(x)$ are defined
as
\begin{equation}
w_{L,R}(x)= \frac{1}{\sqrt{2}}(\psi_1(x)\mp\psi_2(x)).
\end{equation}
The eigenstates $\psi_1(x)$ and $\psi_2(x)$ are orthonormal.
Amplitudes $a_{L,R}$ must satisfy $|a_{L}|^2+|a_{R}|^2 =1$ and their
time derivatives are approximately given by
\begin{equation}\label{wspolczynniki}
    i\dot a_{L,R}(t)= E_0 a_{L,R}(t) -Ka_{R,L}(t)+U_{L,R}|a_{L,R}(t)|^2a_{L,R}(t),
\end{equation}
where
\begin{equation} \nonumber
E_0 = \int w_{L,R}^*(x) \left[-\frac{\partial^2}{\partial x^2} + V(x)
\right] w_{L,R}(x), \,\,\,
K = - \int w_L^*(x) \left[-\frac{\partial^2}{\partial x^2} + V(x)
\right] w_R(x) \, \mbox{d}x,\,\,\, U_{L,R} = \eta \int
|w_{L,R}(x)|^4.
\end{equation}
Note that $E_0$ is the common value of two expressions (for $w_L(x)$
and $w_R(x)$). The
Josephson equations,
\begin{eqnarray} \label{jos1}
\dot{z} &=& - \sqrt{1-z^2} \sin \phi,
\\ \label{jos2}
\dot{\phi} &=& \Lambda z +  \frac{z}{\sqrt{1-z^2}} \cos \phi+\Delta
\end{eqnarray}
will follow upon defining
\begin{eqnarray}
&a_{L,R} =& \sqrt{\frac{1\mp z}{2}} \exp (i
\theta_{L,R}),\,\,\,\,\phi = \theta_L -\theta_R,\nonumber
\\&\Lambda =&
(U_R+U_L)/4K,\,\,\,\,\,\,\,\,\,\,\,\,\,\Delta =
(U_R-U_L)/4K,\nonumber
\end{eqnarray}
and rescaling the time $2Kt \rightarrow t$. Here $\Lambda$ is the
ratio of nonlinear coupling to tunneling and $\Delta$ is the
difference in the depth of the wells. With our simplifications and
substitutions the suitably normalized Hamiltonian of the system can
be obtained in the form (see also \cite{Smerzi,Raghavan})
\begin{equation}
H/K = E_0/K -  \sqrt{1-z^2} \cos \phi +
\frac{\Lambda}{2}(1+z^2)-\Delta z,
\end{equation}
The parameter $\Lambda$ is positive for repulsive interaction ($\eta
>0$) and negative for attractive interaction ($\eta <0$). Note the
two symmetries: $\Lambda \longrightarrow - \Lambda, \,\, \phi
\longrightarrow \phi + \pi, \,\, z \longrightarrow - z$ and $\Delta
\longrightarrow - \Delta, \,\, z \longrightarrow - z, \,\, \phi
\longrightarrow - \phi$. The first of these symmetries implies that
completely solving for $\eta>0$ gives the solution for $\eta<0$.

These equations differ from those governing Josephsonian
oscillations in superconducting junctions by two additional terms:
one proportional to $\Lambda$ which derives from the nonlinear
interaction (it has the same sign as $\eta$) and the constant
$\Delta$, owing its existence to the asymmetry of the potential.

\begin{figure}
\includegraphics[width=8.5cm]{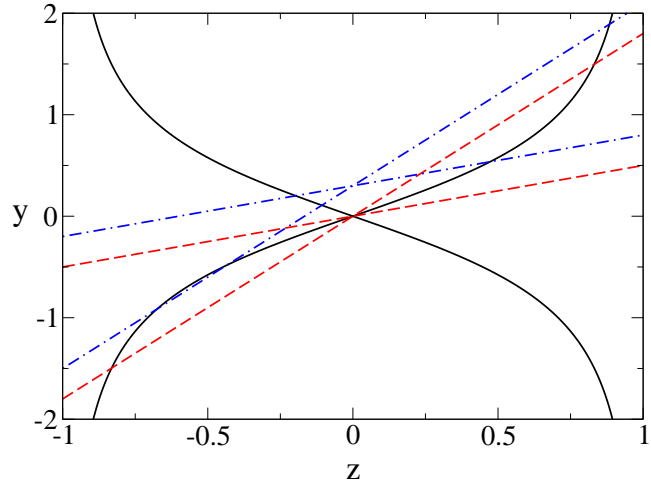}
\caption{(Color online) Stationary solutions of the Josephson equations,
represented by the intersections of the solid lines ($y=\pm
\frac{z}{\sqrt{1-z^2}}$) with the dashed ones ($y=\Lambda z$ for $\Lambda
= 0.5; 1.8$ - symmetric potential case) and dashed dotted line ($y=\Lambda z
+\Delta  $ for $\Lambda =0.5; 1.8$ and $\Delta =0.3$- asymmetric potential case).}
\label{rysnowy}
\end{figure}

Consider the stationary solutions of the Josephson equations. From
Eq.~(\ref{jos1}) we see that $\phi=0; \pm \pi$. In Fig.
\ref{rysnowy} we see how to find them graphically. For $\Delta = 0$
there are always two solutions with $z=0$. The other two solutions
appear for nonzero $z$ when $\Lambda > 1$ (or $\Lambda < -1$). In the
case of nonzero $\Delta$ there are also always at least two solutions. The other
two appear above $\Lambda$ equal $\Lambda_c=(1-\Delta^{2/3})^{3/2}$.

\begin{figure}
\includegraphics[width=8.5cm]{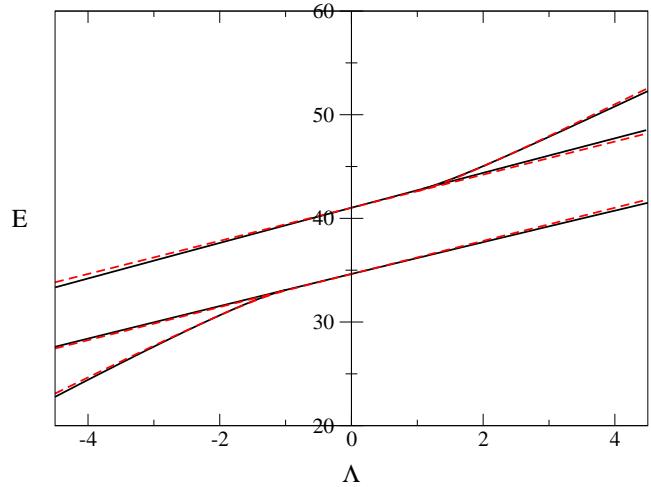}
\caption{(Color online) Bifurcation diagram. The two sets of curves that almost
coincide are obtained from the Josephson Hamiltonian (dashed lines) and the exact
formulas  of sections II and III (solid lines). Beyond the bifurcation point the
symmetric solutions are unstable  and the asymmetric ones are stable.
Each point on the stable bifurcated branch corresponds to two mirror
image solutions.} \label{rys3}
\end{figure}

Having the stationary solutions we can draw the energy dependence on
$\Lambda$, shown in Fig.~\ref{rys3}. The two lowest eigenvectors of the
nonlinear Schr\"{o}dinger equation (solid lines) are compared with
those resulting from the Josephson Junction approach (dashed lines)
for the case of equally deep wells. Notice the good agreement.

\begin{figure}
\includegraphics[width=8.5cm]{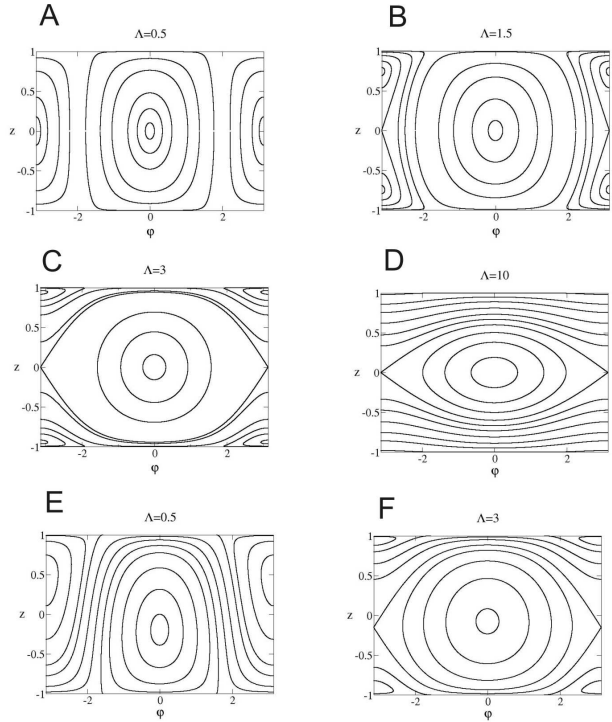}
\caption{Phase space diagrams. The first four frames correspond to
symmetric potential wells, the latter two to a deeper well on the
left ($\Delta =0.3$). Note the differences in the trajectories between the cases B
and E, especially in the "waves" that cover the entire
$\phi$-range. The fixed points are still present in the corners of
the fourth frame (D) but do not turn up on this scale.} \label{rys2}
\end{figure}

Now consider the dynamics of the $\Delta=0$ case. Constant energy
contours in $z$, $\phi$ phase space followed by the system for
positive $\Lambda$ are shown in Fig. \ref{rys2}, (A-D), one each for
$\Lambda < 1$, $1<\Lambda <2$ and two for $\Lambda >2$, each of
which is generic. The difference between the second and third case
concerns the possibility of self trapping solutions oscillating
about an average $z$ such that $\phi$ covers all possible values in
the third case. However, the fixed point dynamics is common to the
latter three cases. Fixed points are at: (1) $z=0$, $\phi=0$; (2)
$z=0$, $\phi=\pm \pi$; (3) $z=\sqrt{1-\Lambda^{-2}}$, $\phi =\pm
\pi$, $\Lambda >1$. The latter pair bifurcate from the second point
as we increase $\Lambda$ through $\Lambda =1$, see Fig.~3 for an
illustration of how this happens.

\begin{figure}
\includegraphics[width=8.5cm]{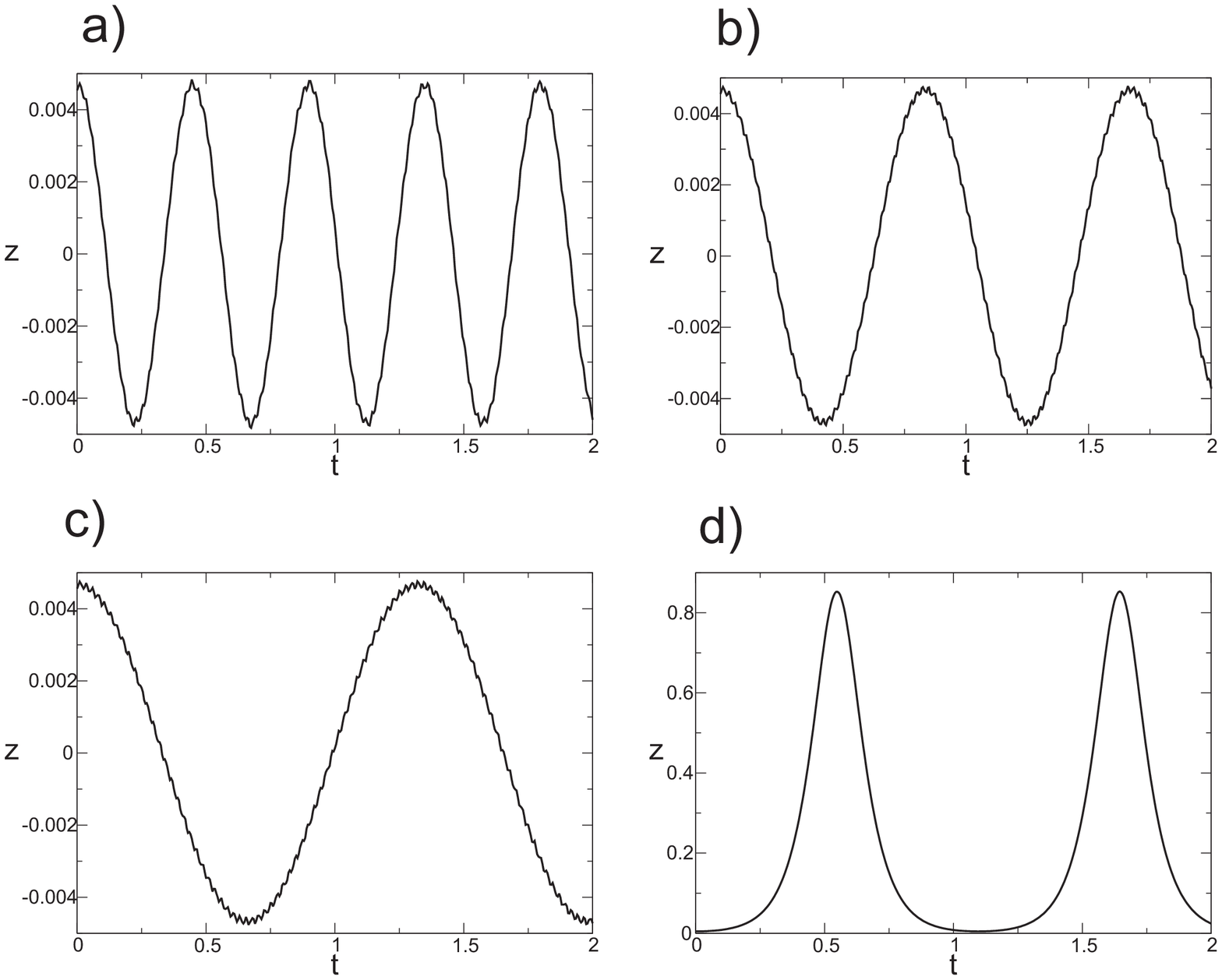}
\caption{Evolution of $z(t)$ for the antisymmetric states, $\phi = \pi$: a) $\eta
= -10$ , $\Lambda = -4.52$ ; b) $\eta = -1$, $\Lambda = -0.452$ ; c)
$ \eta = 1$, $\Lambda = 0.452$; d) $\eta=10$, $\Lambda = 4.52$. As
we see in the case of a), b) and c) the solution is stable. The
periods of oscillations match the formulas derived from the values
of $\lambda$ given in Section 3, $T=2\pi/|\lambda|$. In case d)
the solution is unstable. The stability of the asymmetric bifurcated branch has also been confirmed.} \label{rys4}
\end{figure}

We will now look at the stability of the three classes of fixed
points. Assume perturbations such that $z \rightarrow z+\delta z
e^{\lambda t}$ and $\phi \rightarrow \phi+\delta \phi e^{\lambda
t}$. Simple calculations give values of $\lambda$ for the three
categories: \\
(1) $\lambda^2 = -(1+\Lambda)$ (phase point in the ($\phi$,$z$) plane moves
on an ellipse like trajectory around ($0$,$0$)) \\
(2) $\lambda^2 = \Lambda -1$ (fixed point stable when $\Lambda < 1$,
but when $\Lambda > 1$, any perturbation moves out along an arm of
the separatrix emerging from ($\pm \pi$,$0$)) \\
(3)  $\lambda^2 = 1-\Lambda^2 $  (phase point moves on an ellipse
like trajectory around one of
($\pm \pi$,$\pm \sqrt{1-\Lambda^{-2}}$)) \\
Thus, according to this criterion the first fixed point is always
stable for $\Lambda >-1$. The antisymmetric solution (2) is stable
for $\Lambda <1$ and unstable for $\Lambda>1$. The bifurcated pair
(3) is always stable and we have a typical pitchfork bifurcation at
$\Lambda =1$. These results are in full agreement with a numerical stability
analysis based on the nonlinear Schr\"{o}dinger equation (see
Fig.~\ref{rys4}). We might add that they contradict some statements
in the literature, e.g. \cite{Dagosta} and \cite{Weinstein}.

One might wonder how the Josephson bifurcation picture ties up with the exact
solutions considered earlier. Suppose we have an analytic solution
given by $m_1$, $m_2$, $m_3$, $\mu$ and $d$. As $\phi=\pm \pi$ in our
considerations, this solution clearly corresponds to one of the
bifurcated fixed points in the Josephson model. How can we determine
the corresponding value of $\Lambda$ so that $z$ takes the proper
value? A good approximation to $z$ when $d$ is small, is (Appendix):
\begin{equation}
z = \frac{\left[(k_1^2-k_3^2)\omega - k_1 E(k_1\omega |m_1) +k_3
E(k_3\omega |m_3)\right]}{\left[(k_1^2+k_3^2)\omega - k_1
E(k_1\omega |m_1) - k_3 E(k_3\omega |m_3)\right]}
\end{equation}
and $\Lambda = (1-z^2)^{-1/2}$. The apparent independence of $d$ is
deceptive, as $d$ will determine $m_1$ and $m_3$, above. In
particular when $d=0$, $m_1=m_3$, $k_1=k_3$ and $z=0$ as expected.
The antisymmetric solution is recovered.

When $\Lambda\neq 0$, the fixed points are still located at
$\phi=0;\pm \pi$. The stationary  values of $z$ are now roots of the quartic
\begin{equation}
\Lambda^2 z^4 + 2\Delta \Lambda z^3 + (\Delta^2+1-\Lambda^2) z^2 -
2\Lambda \Delta z - \Delta^2=0
\end{equation}
and are shifted down as compared to the case of $\Delta=0$.
Illustrations of how phase diagrams are modified as compared to the
symmetric potential case are given in Fig.~\ref{rys2}, E and F. Now phase curves covering all possible $\phi$ values and
such that $z$ changes sign are possible. As
we increase $\Lambda$ from zero, a new double fixed point suddenly
appears at a critical value of $\Lambda$ equal
$\Lambda_c=(1-\Delta^{2/3})^{3/2}$, and bifurcates as we increase
$\Lambda$. Thus for $\Lambda > \Lambda_c $ we have two new fixed
points.  A stability analysis yields
results similar to the above, for $\Delta=0$, but values of
$\lambda_i$ are now given in terms of roots of the quartic $z_i$.

\section{Conclusions}

In this paper we have thoroughly investigated the behavior of
Bose-Einstein condensates in double square well potentials, both
symmetric and asymmetric. A simple method for obtaining exact
solutions for repulsive interaction was outlined (similarly as in
\cite{Zin} for attractive interaction). We treat the system both
exactly and by a Josephson Junction model. We have checked the
Josephson model results, both static and dynamic, against exact
calculations. Agreement is suprisingly good. Some controversies
about the  stability, to be found in the literature, have been
resolved.

\section{Acknowledgement}

The authors would like to acknowledge support from KBN Grant 2P03
B4325 (E.I.),  KBN Grant 1P03B14629 (P.Z.)  and the Polish Ministry of Scientific Research and
Information Technology under grant PBZ-MIN-008/P03/2003 (M.T.).

Consultations with Professor George Rowlands were very helpful.

\section{Appendix}

\subsection{Symmetry preserving case (symmetric wells)}

To find the $\Delta \mu$ correction we eliminate $A$ and $A_2$ from
equations (\ref{8}) and (\ref{9})
\begin{eqnarray} \label{rF}
F=\frac{k
\mbox{cn}(k\omega|m)\mbox{dn}(k\omega|m)}{\mbox{sn}(k\omega|m)}+
\frac{k_2\mbox{dn}(k_2b|m_2)}{\mbox{sn}(k_2b|m_2)\mbox{cn}(k_2b|m_2)}=0.
\end{eqnarray}
and calculate the perfect differential of $F(m,m_2,\mu)$ for small
$\eta$. As the first two differentials follow from
Eqs.~(\ref{eq6}) $\Delta \mu$ can be so obtained. This
calculation is somewhat less straightforward. It completes the
calculation of $m$, $m_2$ and $\mu$ in the small $\eta$ limit, our
starting point.

In the limit $m$ and $1-m_2$ tending to zero Eq.~(\ref{eq7}) is
recovered from (\ref{rF}). The general equation for small increments
of $m, m_2$ and $\mu$ is
\begin{eqnarray}
\Delta F=\frac{\partial F}{\partial m}\Delta m  +\frac{\partial
F}{\partial m_2}\Delta m_2 + \frac{\partial
F}{\partial \mu}\Delta \mu =0 \nonumber \\
\Delta m = m, \,\,\,\,  \Delta m_2 = m_2-1, \,\,\,\, \Delta \mu =
\mu- \mu_0,
\end{eqnarray}
and so
\begin{equation}\label{pop}
\Delta \mu =\left(-\frac{\partial F}{\partial m} m +\frac{\partial
F}{\partial m_2}(1- m_2)\right) \left(\frac{\partial F}{\partial
\mu}\right)^{-1},
\end{equation}
where in the perturbation limit $m$ and $1-m_2$ are proportional to
$\eta$ and are given by equation (\ref{eq6}). We find after some
calculations using known identities \cite{Abram}
\begin{eqnarray}
\frac{\partial F}{\partial m} & = &- k \left(
\frac{3}{4}S+\frac{1}{4}\sin 2k\omega \right)
\nonumber \\
\frac{\partial F}{\partial m_2} & = &
k_2\left(\frac{3}{4}L-\frac{1}{4}\sinh(2k_2b) \right)
\nonumber \\
\frac{\partial F}{\partial \mu} & = &
\frac{1}{2k}S-\frac{1}{2k_2}L=-\frac{A^{-2}} {2\sin^2 k\omega}
\end{eqnarray}
and
\begin{eqnarray}
S & = & \cot(k\omega)-\frac{k\omega}{\sin^2 (k\omega)}
\nonumber \\
L & = & \coth (k_2b)- \frac{k_2b}{\sinh^2 (k_2b)},
\end{eqnarray}
where $k$ and $k_2$ are taken in the linear limit.

\subsection{Symmetry breaking case (symmetric wells)}

We obtain
\begin{eqnarray}
A_1^2 = \frac{2m_1k_1^2}{\eta}, \ \ \ \ \ A_3^2 =
\frac{2m_3k_3^2}{\eta}, \ \ \ \ \ \   A_2^2 =
\frac{2(1-m_2)k_2^2}{\eta}, \nonumber \\
 \mu = (1+m_1)k_1^2 =
(1+m_3)k_3^2 = (m_2 - 2)k_2^2 + V_0.\label{eq88}
\end{eqnarray}
The continuity conditions at $x=\pm b$ are now generalized to:
\begin{eqnarray}\label{eq8}
g_0 &=& k_1\sqrt{m_1}\,\,
\mbox{sn}(k_1\omega|m_1)-k_2\sqrt{(1-m_2)}\,\,
\mbox{sc}(k_2(b-d)|m_2)=0   \\
g_1 &=& k_3\sqrt{m_3}\,\,
\mbox{sn}(k_3\omega|m_3)-k_2\sqrt{(1-m_2)}\,\,
\mbox{sc}(k_2(b+d)|m_2)=0  \nonumber \\
g_2 &=&  k_1^2\sqrt{m_1}\,\,
\mbox{cn}(k_1\omega|m_1)\,\,\mbox{dn}(k_1\omega|m_1)
+k_2^2\sqrt{1-m_2}\,\,
\mbox{dc}(k_2(b-d)|m_2)\,\,\mbox{nc}(k_2(b-d)|m_2)=0 \nonumber \\
g_3 &=& k_3^2\sqrt{m_3}\,\,
\mbox{cn}(k_3\omega|m_3)\,\,\mbox{dn}(k_3\omega|m_3)
+k_2^2\sqrt{1-m_2}\,\,
\mbox{dc}(k_2(b+d)|m_2)\,\,\mbox{nc}(k_2(b+d)|m_2)=0 \nonumber.
\end{eqnarray}
and the normalization condition is now:
\begin{eqnarray}\label{g4row}
g_4 &=& {2k_2} \left[ \mbox{sn}(k_2(b-d)|m_2) \,
\mbox{dc}(k_2(b-d)|m_2) + \mbox{sn}(k_2(b+d)|m_2) \,
\mbox{dc}(k_2(b+d)|m_2) \right]
\\ \nonumber
& &- {2k_2} \left[ E(k_2 (b+d)|m_2) +E(k_2 (b-d)|m_2)\right]
\\ \nonumber
& &+ {2k_1}\left[ k_1\omega -
 E(k_1 \omega |m_1)\right]
+ {2k_3}\left[ k_3\omega -
 E(k_3 \omega |m_3) \right] = {\eta}.
\end{eqnarray}

If we can assume $k_2d$ much smaller than one, Eqs. (\ref{eq8}) and
(\ref{g4row}) up to second order simplify to:
\begin{eqnarray}
& & \chi(m_1) - \chi(m_3) = 2 D_2 d \frac{(2-m_2)\mbox{sc}(k_2b|m_2)
+ 2(1-m_2) \mbox{sc}^3(k_2b|m_2)}{m_2}
\\ \nonumber
\\ \nonumber
& & \chi(m_1)+\chi(m_3) = -\frac{2D_2}{k_2} \mbox{dc}(k_2b|m_2)
\mbox{nc}(k_2b|m_2)
\\ \nonumber
\\ \nonumber
& & \psi(m_1) - \psi(m_3) = -2d
k_2^2\sqrt{1-m_2}\mbox{dc}(k_2b|m_2)\mbox{nc}(k_2b|m_2)
\\ \nonumber
\\ \nonumber
& & \psi(m_1) + \psi(m_3) = 2k_2 \sqrt{1-m_2}\mbox{sc}(k_2b|m_2)
\\ \nonumber
\\ \nonumber
& & \eta =  \left[ \phi(m_1) + \phi(m_3) \right] + 4k_2
\left[\mbox{sn}(k_2b|m_2)\mbox{dc}(k_2b|m_2)-E(k_2b|m_2) \right]
\end{eqnarray}
where:
\begin{eqnarray} \nonumber
\chi(m) &=& \frac{\mu}{1+m} \sqrt{m}\,\mbox{cn}(k\omega |m)\,
\mbox{dn}(k\omega |m)
\\ \nonumber
\psi(m) &=& \sqrt{\frac{\mu}{1+m}} \sqrt{m}\,\,{\mbox{sn}(k\omega
|m)}
\\ \nonumber
D_2 &=& \left[\frac{V_0-\mu}{2-m_2}\right]^{3/2} \sqrt{1-m_2} \ \ \
\ k^2 = \frac{\mu}{1+m} \ \ \  k_2^2 = \frac{V_0-\mu}{2-m_2}
\\ \nonumber
\phi(m) &=& 2k \left[ k\omega - E(k\omega |m) \right],
\end{eqnarray}
and $z=\left[ \phi(m_1) - \phi(m_3) \right] / \left[ \phi(m_1) + \phi(m_3) \right]$. By comparing
$d$ determined by the first and third equations we can reduce the
system to just four equations for four unknowns, $m_1$, $m_2$, $m_3$
and $\mu$.

\end{document}